\documentclass[aip,jcp,reprint]{revtex4-2}
\usepackage{graphicx,amssymb,amsmath}
\usepackage{mhchem}
\usepackage{color}
\usepackage[dvipsnames]{xcolor}
\usepackage{tikz}
\usepackage{circuitikz}
\usepackage{physics}
\begin{document}
\title{Ubiquitous preferential water adsorption to electrodes in water/1-propanol mixtures detected by electrochemical impedance spectroscopy}
\author{Haruto Iwasaki$^1$, Yasuyuki Kimura$^1$, and Yuki Uematsu$^{2,3}$}
\email[]{uematsu@phys.kyutech.ac.jp}
\affiliation{Department of Physics, Kyushu University, Motooka 744, Fukuoka 819-0395, Japan}
\affiliation{Department of Physics and Information Technology, Kyushu Institute of Technology, Iizuka 820-8502, Japan}
\affiliation{PRESTO, Japan Science and Technology Agency, 4-1-8 Honcho, Kawaguchi, Saitama 332-0012, Japan}
\date{\today}

\begin{abstract}
The electric double layer is an important structure that appears at charged liquid interfaces, and it determines the performance of various electrochemical devices such as supercapacitors and electrokinetic energy converters.
Here the double-layer capacitance of the interface between aluminum electrodes and water/1-propanol electrolyte solutions is investigated using electrochemical impedance spectroscopy. 
The double-layer capacitances of mixture solvents are almost the same as those of water-only electrolyte solutions, and the double-layer capacitance of 1-propanol-only solutions are significantly smaller than those of other volume fractions of water.
The qualitative variation of the double-layer capacitances with the water volume fraction is independent of the electrolyte types and their concentrations.
Therefore, these results can be explained by ubiquitous preferential water adsorption caused by the hydrophilicity of the electrode surface.
\end{abstract}


\maketitle

\section{Introduction}

The structure of the electric double layer is important in various applications, such as supercapacitors \cite{Bozym_2015} and electrokinetic energy conversion \cite{Siria_2013}.
Thus, aqueous electric double layers have been intensely studied theoretically and experimentally \cite{Gouy_1910, Chapman_1913, Stern_1924,Grahame_1947,Valette_1981,Pajkossy_1996,Lamperski_1996, Damaskin_2011, Bonthuis_2012, Uematsu_2018_1, Uematsu_2018_2, Uematsu_2021}.
The standard model of an aqueous electric double layer called Stern-Gouy-Chapman theory considers two layers at the solid/solution interface \cite{Stern_1924, Gouy_1910, Chapman_1913}.
The first layer is the Stern layer (or Helmholtz layer) whose thickness is the molecular size.
The Stern layer consists of mono- or a few layers of water molecules, and in this layer, the dielectric and viscous properties are much different from those of bulk \cite{Bonthuis_2012, Uematsu_2018_1}. 
The second layer is the diffuse layer, in which the electrostatic and entropic force exerted on ions are balanced. 
As a result, this layer is charged, and the thickness, which is called Debye length, depends on the salt concentration.
The Stern-Gouy-Chapman theory derives the total capacitance of the electric double layer as a series circuit that consists of the two capacitors of the Stern layer and diffuse layer, respectively.
This picture has succeeded in explaining and predicting the double-layer properties, for example, the differential capacitance of mercury electrodes \cite{Grahame_1947}, and the zeta potentials of various metal oxides and hydrophobic materials \cite{Uematsu_2018_1}.
However, the Stern-Gouy-Chapman theory cannot explain the recent experimental data on the double-layer capacitances of polycrystal platinum or indium tin oxide electrodes \cite{Aoki_2013, Khademi_2020, Aoki_2021}, although these studies did not seriously take into account the variation of the surface potential with the salt concentration.

When the solvent is altered by organic solvents or their aqueous mixture \cite{Izutsu_2002}, additional effects on the double-layer capacitance are necessary to consider such as ionic association \cite{Holovko_2001}, preferential solvation of ions \cite{Yabunaka_2017}, preferential adsorption of solvents to the interface \cite{Aoki_2018}, and phase separation in bulk phase \cite{Cruz_2018,Cruz_2019}. 
Thus, the structure of double layers in organic solvents \cite{Kim_2010,Hou_2014,Bozym_2015} and their aqueous mixtures \cite{Srivastava_1980, Hidalgo_Alvarez_1985, Schwer_1991, Rubio_Hern_ndez_1998,  Yabunaka_2017, Aoki_2018, Thanh_2019} are different from that of aqueous ones, and the theoretical description of it is still insufficient. 
Nevertheless, electric double layers of organic solvents and their aqueous mixtures are significantly important because various chemical and electrochemical products use organic solvents such as detergent, paint, ink, and batteries.
One of the advantages to employ binary mixture as solvents is that the bulk properties such as dielectric constant, viscosity, and conductivity are continuously adjustable by varying the composition of the mixture solvent.  
Therefore, it is necessary to understand how the interfacial properties of the electric double layer in aqueous mixture solvents are modified as a function of the solvent composition.

In this context, the zeta potential of silica in contact with water/organic liquid mixtures was first studied by Kenndler \cite{Schwer_1991} {\it et al.}
The organic liquids were methanol, ethanol, 2-propanol, acetonitrile, acetone, and dimethyl sulfoxide, and they all are soluble in water with $10\,$mM potassium chloride. 
The zeta potentials were calculated from the measured electro-osmotic mobility, the dielectric constant, and the viscosity of the mixture solvents using Smoluchowski equation. 
The obtained zeta potentials as a function of the molar fraction of the organic liquids revealed that all of the organic liquids linearly decrease the magnitude of zeta potentials except for acetone \cite{Schwer_1991}.
In that study, the linear decrease of the absolute zeta potential by added organic liquids was attributed to the reduction of the Stern-layer capacitance and/or the suppression of surface charge density caused by the adsorption of organic molecules on the surface. 
Besides the zeta potential, the composition dependence of the double-layer capacitance in water/acetonitrile mixture with $100\,$mM salt concentration was studied by Aoki {\it et al.} \cite{Aoki_2018} 
The double-layer capacitance of polycrystal platinum electrode/solution interface exhibits characteristic variation in the range of water molar fraction from 0 to 0.2, and it remains constant in the range from 0.2 to 1.
This result suggests that the double-layer structure of aqueous mixture solvents is quite similar to that in water, and the effect of added organic liquids  is not significant for those in intermediate water fractions.
Assuming that the double-layer capacitance is dominated by the capacitance of the Stern layer, preferential adsorption of water makes the double-layer capacitance of the aqueous mixture solvent almost the same value as that of water-only electrolyte solutions \cite{Aoki_2018}.
However, this picture is not consistent with the previous study on the silica surface in aqueous mixtures \cite{Schwer_1991}, because, in that study, the magnitude of zeta potentials was linearly decreased by small addition of the organic liquids into water, suggesting no preferential water adsorption to the interface. 
Thus, it is still unclear whether the preferential water adsorption to electrodes in aqueous mixture solvents is ubiquitous or specific to the interface between polycrystal platinum electrodes and water/acetonitrile mixture. 
Furthermore, the mechanism of preferential water adsorption was attributed to the difference of the solvent-surface interaction between water and organic solvents.
Although the ion-specific effect on the double-layer capacitance in water-only or organic-solvent-only electrolyte solutions has been reported \cite{Grahame_1947, Han_2018}, the role of ions in preferential water adsorption in aqueous mixture solvents is not clear. 
Therefore, it is necessary to study the preferential water adsorption to electrodes with many types and concentrations of electrolytes.

In this paper, we study the double-layer capacitance of electrode/aqueous mixture solvent, and we focus on preferential water adsorption to electrodes in a mixture of water and 1-propanol.
To clarify the origin of preferential water adsorption, we used four kinds of electrolytes and varied their concentrations to observe the ion-specific effect on the double-layer capacitance. 
The examination of many types and concentrations of electrolytes has never done before in the study of the double-layer capacitance in aqueous mixture solvents to the best of our knowledge. 
The experimental results demonstrated that preferential water adsorption was almost independent of ion species and their concentrations.
This suggests that the main origin of the preferential adsorption is the interaction between the electrode and the solvent molecules, and the ions play a minor role in preferential water adsorption.
Because our system uses aluminum electrodes and 1-propanol as an organic liquid, which are different from platinum electrodes and acetonitrile in the previous study \cite{Aoki_2018}, preferential water adsorption is expected to be ubiquitous to broad types of water/organic liquid mixture, metallic electrodes, and electrolytes. 

\section{Methods}

\subsection{Experimental methods}

A commercial electroporation cuvette (Biorad, 1652086) is used as an cell for electrochemical impedance spectroscopy. 
The electrodes are made of aluminum, and the gap is $2\,$mm. 
The cross-section of the electrodes is approximately $S=10\,$mm$\,\times\, 20\,$mm.
The electrochemical cell was soaked in neutral detergent solutions ($1\,$v\%, Cica Clean LX-II, Kanto Chemical) for at least 12 hours, rinsed with deionized water, and cleaned with an ultrasonic bath for ten minutes. 
The sample liquid, $1\,$mL, was injected into the electrochemical cells by a micropipette after prewashing with the same sample liquid at least three times.
The electrochemical impedance spectroscopy was performed by a potentiostat (PalmSens, PalmSens4). 
Because two-electrode measurements were performed, the lead for the counter electrode was connected to the lead for the reference electrode.
The impedance measurement was performed at the open circuit potential and room temperature, and the amplitude of the voltage was $10\,$mV.
The frequency range was from $1\,$Hz to $10^6\,$Hz.

The solvents used were 1-propanol (Fujifilm Wako, 99.5\%) and ultrapure water produced from a water purification system (Merk Millipore, Direct-Q 5UV).
1-propanol was dried using molecular sieves (Fujifilm Wako, 3A 1/16) for at least 24 hours. 
The sodium chloride (NaCl, Fujifilm Wako, 99.5\%), sodium percolate (NaClO$_4$ Fujifilm Wako, 95.0+\%), tetrabutylammonium percolate (TBAClO$_4$, Tokyo Kasei, 98.0\%), and tetrabutylammonium chloride (TBAC, Tokyo Kasei, 98.0\%) were used as received without further purification.
Aqueous mixture solutions of NaCl, TBAC, and NaClO$_4$ were prepared by making a dense water-only electrolyte solution and diluting it with water and 1-propanol to achieve a given volume fraction and salt concentration. 
Aqueous mixture solutions of TBAClO$_4$ were prepared by making a dense 1-propanol-only electrolyte solution and diluting it.
Because the salt concentration is small, we neglected the volume of salts and the volume change caused by the mixing of solvents.  

The reason to select 1-propanol as an organic solvent is that it is soluble with water in any ratio, whereas it has a tendency of phase separation when a large amount of salt is added \cite{Gomis_1994}. 
The reason to select these four types of electrolytes is that they are pairs of typical hydrophilic (sodium and chloride ions) and hydrophobic ions (tetrabutylammonium and percolate ions) \cite{Marcus_2015}.

\subsection{Data analysis}

\begin{figure}[t]
\includegraphics{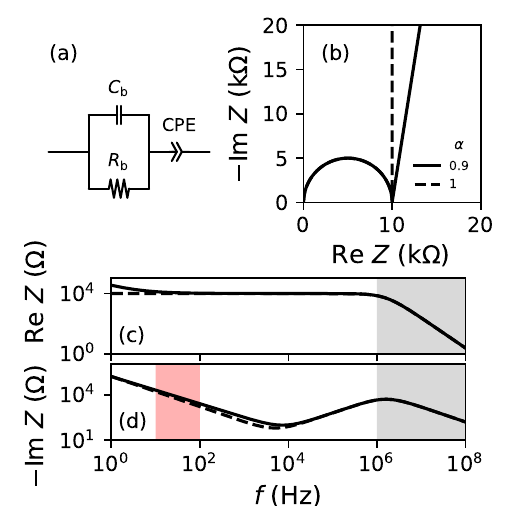}
\caption{
(a) Equivalent circuit model of this study. $C_\mathrm{b}$ denotes the bulk capacitance of the solution whereas $R_\mathrm{b}$ denotes the bulk resistance of the solution.
CPE denotes a constant phase element of an electric double layer.
(b) Nyquist plot of the equivalent circuit model.
(c) $\Re Z$ as a function of frequency.
(d) $\Im Z$ as a function of frequency.
The parameters of the elements are $R_\mathrm{b}=10^4\,\Omega$, $C_\mathrm{b}=10^{-11}\,$F, $C_\mathrm{d}=1\,\mu$F/cm$^2$, $S=2\,$cm$^2$, and $\tau_0 = (1/2\pi)\,$s.
The black solid lines denote $\alpha=0.9$ whereas the broken lines denote $\alpha=1$. 
The shaded regions with gray ($10^6\,$Hz$\,< f\le 10^8\,$Hz) in (c) and (d) denote the out of range in our measurements, whereas the shaded region with red ($10^1\,$Hz$\,\le f \le 10^2\,$Hz) in (d) denotes the fit range of eq.~\ref{eq:imz}.
}
\label{fig:1}
\end{figure}

In this section, we briefly describe the way of extracting double-layer capacitance from the impedance spectroscopy data.
Normally the complex capacitance spectrum is defined as $C(\omega)=1/i\omega Z$ where $Z$ is the impedance, $i$ is the imaginary unit, and $\omega$ is the angular frequency.\cite{Pajkossy_1996}
Thus, the low-frequency limit of $\Re C$ is the double-layer capacitance.
However, the obtained double-layer capacitances often have frequency dependences except for single-crystal electrodes with nonabsorbing ions \cite{Valette_1981,Pajkossy_1996}, and it has been often modeled by the constant phase element (CPE), of which impedance is given by $Z_\mathrm{CPE}\sim (i\omega)^{-\alpha}$, where $\alpha\quad(0<\alpha\le 1)$ is an adjustable parameter.
The frequency dependence of the double-layer capacitance has been explained not only by geometrical roughness but also by energetic inhomogeneity such as the facet of the crystals or the ion adsorption properties \cite{Liu_1985,Pajkossy_1994}.
To interpret the CPE as a double-layer capacitance, an arbitrary frequency is often introduced \cite{Brug_1984,Orazem_2013,Khademi_2020,Aoki_2018}.
Therefore, a quantitative comparison between the Stern-Gouy-Chapman theory and the experimental double-layer capacitances is not straightforward.

The impedance of the equivalent circuit model in our study is shown in Fig.~\ref{fig:1}a, and its impedance is given by
\begin{equation}
Z = \left(\frac{1}{R_\mathrm{b}}+i\omega C_\mathrm{b}\right)^{-1}+2Z_\mathrm{CPE},
\label{eq:1}
\end{equation}
where $R_\mathrm{b}$ is the solution resistance, $C_\mathrm{b}$ is the solution capacitance, and the factor $2$ in front of $Z_\mathrm{CPE}$ means the two electrodes in a series circuit. 
The impedance of CPE, $Z_\mathrm{CPE}$, is defined as
\begin{equation}
Z_\mathrm{CPE}=\frac{\tau_0}{(i\omega \tau_0)^{\alpha}C_\mathrm{d}S},
\end{equation}
where $\tau_0$ is the arbitrary time, $S$ is the electrode surface area, and $C_\mathrm{d}$ is the double-layer capacitance per area.
Normally, the CPE impedance is defined as $Z_\mathrm{CPE}=(i\omega)^{-\alpha}/Q_0$, where $Q_0$ is the effective capacitance.
However, in this definition, the unit of $Q_0$ is not farad, and quantitative discussion of the double-layer capacitance is difficult. 
Thus, we introduce an additional time scale, $\tau_0=(1/2\pi)\,$s.
This is similar to the previous studies \cite{Hou_2014, Aoki_2018}, whereas some studies used a different time scale $\tau_0=C_\mathrm{d}R_\mathrm{b}$. \cite{Brug_1984,Orazem_2013,Khademi_2020}.

The impedance spectroscopy provides $\Re Z$ and $-\Im Z$, which can be calculated from eq.~\ref{eq:1} as
\begin{eqnarray}
\Re Z &=& \frac{1/R_\mathrm{b}}{(1/R_\mathrm{b})^2+(\omega C_\mathrm{b})^2}+\frac{2\tau_0\cos(\pi\alpha/2)}{C_\mathrm{d}S(\omega\tau_0)^\alpha},\\ 
-\Im Z &=& \frac{\omega C_\mathrm{b}}{(1/R_\mathrm{b})^2+(\omega C_\mathrm{b})^2}+\frac{2\tau_0\sin(\pi\alpha/2)}{C_\mathrm{d}S(\omega\tau_0)^\alpha}. \label{eq:4}
\end{eqnarray}
Fig.~\ref{fig:1}b is the Nyquist plot with typical parameters.
The parameters of the elements are $R_\mathrm{b}=10^4\,\Omega$, $C_\mathrm{b}=10^{-11}\,$F, $C_\mathrm{d}=1\,\mu$F/cm$^2$, $S=2.0\,$cm$^2$, and $\tau_0 = (1/2\pi)\,$s.
These parameters are close to those obtained in our experiments, whereas the frequency range is limited from $10^0\,$Hz to $10^6\,$Hz in our experiments. 
The exponent $\alpha$ is varied as $\alpha=1$ for the broken lines and $\alpha=0.9$ for the solid lines.
The Nyquist plot is composed of a semicircle and a straight line. 
The semicircle corresponds to the relaxation caused by the parallel circuit of $R_\mathrm{b}$ and $C_\mathrm{b}$, whereas the straight line corresponds to the double-layer capacitance. 
The angle of the straight line is $\pi/2$ when the double-layer capacitance is ideal ($\alpha=1$), but it tilts with an angle $\pi\alpha/2$ when the element is CPE with $0<\alpha<1$.

Fig.~\ref{fig:1} (c) and (d) are the spectra of real and imaginary parts of the impedance. 
The shaded region with gray ($10^6\,$Hz$\,< f\le 10^8\,$Hz) denotes the out-of-range in our experiments.
$\Re Z$ with $\alpha=1$ exhibits a plateau in low frequency corresponding to the bulk conductance $1/R_\mathrm{b}$ whereas in high frequency the relaxation is caused by the bulk capacitance $C_\mathrm{b}$.
In the case of $\alpha<1$, the frequency dependence of $\Re Z$ appears in the low-frequency region.
The spectrum of $-\Im Z$ exhibits a broad peak in the high-frequency region caused by the relaxation of the parallel $R_\mathrm{b}C_\mathrm{b}$ circuit.
In the low-frequency region, the characteristic downward slope is caused by the double-layer capacitance. 
The slope can be given by
\begin{equation}
-\Im Z = \frac{2\tau_0}{C_\mathrm{d}S}\sin\frac{\pi\alpha}{2}(\omega\tau_0)^{-\alpha},
\label{eq:imz}
\end{equation}
when the first term in eq.~\ref{eq:4} is much smaller than the second term. 
We use eq.~\ref{eq:imz} for extracting $C_\mathrm{d}$ and $\alpha$.
The shaded region with red ($10^1\,$Hz$\,\le f \le 10^2\,$Hz) in Fig.~\ref{fig:1}d denotes the fit range of eq.~\ref{eq:imz}.

\section{Results and discussions}

\begin{figure}[t]
\includegraphics{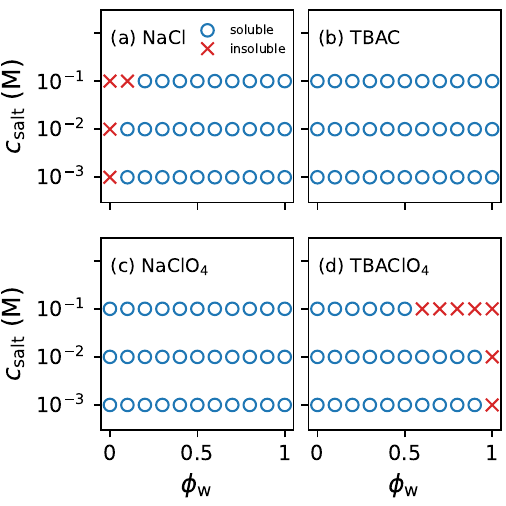}
\caption{
Solubility of four types of electrolytes in a water/1-propanol mixture at room temperature as a function of the salt concentration, $c_\mathrm{salt}$, and the volume fraction of water, $\phi_\mathrm{w}$.
(a) NaCl, (b) TBAC, (c) NaClO$_4$, and (d) TBAClO$_4$. 
The blue open circles denote that the electrolytes are soluble, whereas the red cross marks denote insoluble.
}
\label{fig:27}
\end{figure}

\begin{figure*}[t]
\includegraphics{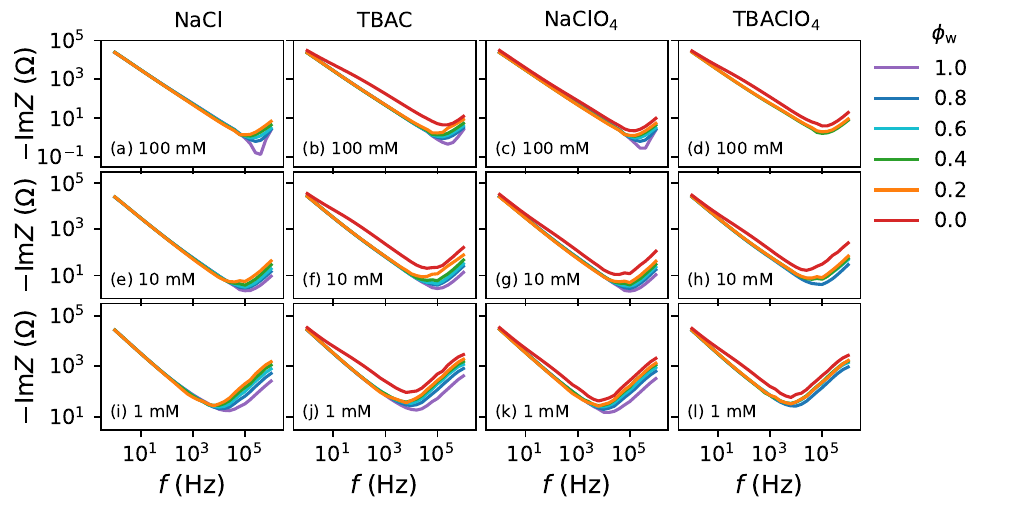}
\caption{
Table of the spectra of $-\Im Z$ with varied electrolytes, salt concentration, and the volume fraction of water.
The color of the lines denotes the volume fraction of water as $\phi_\mathrm{w}=1.0$ (purple), $0.8$ (blue), $0.6$ (light blue), $0.4$ (green), $0.2$ (orange), and $0.0$ (red).
The different columns correspond to the different salt types as NaCl, TBAC, NaClO$_4$, and TBAClO$_4$, whereas the different rows correspond to the different salt concentration as $100\,$mM, $10\,$mM, and $1\,$mM.
The spectra are the average of measurements using four different cells. 
}
\label{fig:49}
\end{figure*}

\subsection{Solubility and Phase Behavior}

We first examined the solubility of four types of electrolytes: NaCl, TBAC, NaClO$_4$, and TBAClO$_4$ in a water/1-propanol mixture.
Fig.~\ref{fig:27} the shows solubility of a given electrolyte at room temperature as a function of the electrolyte concentration and the water volume fraction, and open blue circles denote soluble whereas red cross marks denote insoluble.
In (a), NaCl in water/1-propanol mixtures is partly insoluble in 1-propanol-only and -rich solutions.
In contrast, TBAClO$_4$ in water/1-propanol mixtures is partly insoluble in water-only and -rich solutions as shown in Fig.~\ref{fig:27}d. 
This symmetric behavior of NaCl and TBAClO$_4$ can be explained that NaCl is composed of hydrophilic cations and anions, whereas the TBAClO$_4$ is composed of hydrophobic cations and anions. 
Fig.~\ref{fig:27} (b) and (c) show that TBAC and NaClO$_4$ can be dissolved in any water volume fraction and salt concentrations. 
This can be related to the fact that both electrolytes are composed of a pair of hydrophilic (sodium or chloride ions) and hydrophobic ions (tetrabutylammonium or percolate ions). 

\subsection{Double layer capacitance}

\begin{figure}[t]
\includegraphics{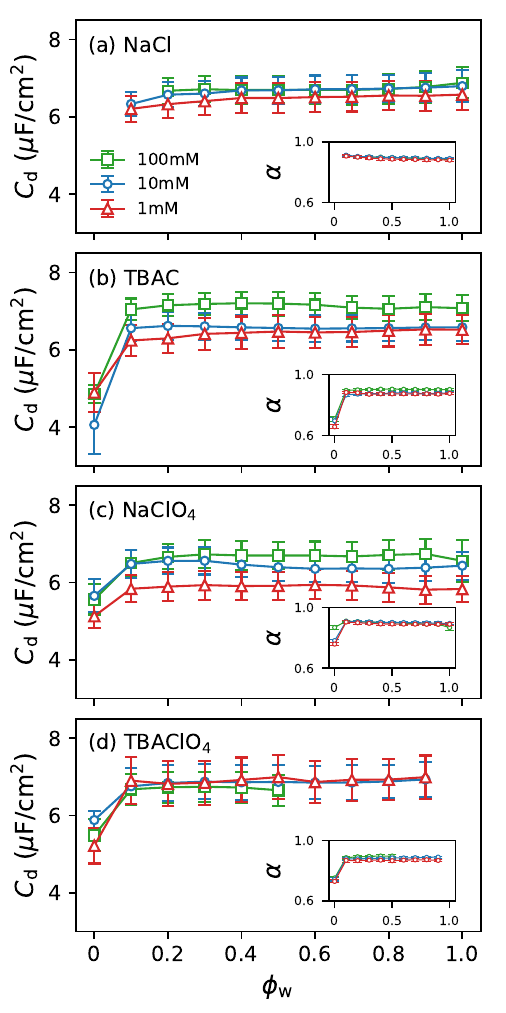}
\caption{
Double-layer capacitance as a function of the water volume fraction.
We used $S=2.0\,$cm$^2$ and $\tau_0=(1/2\pi)\,$s.
(a) NaCl, (b), TBAC, (c) NaClO$_4$, and (d) TBAClO$_4$.
The green squares are $100\,$mM, the blue circles are $10\,$mM, and the red triangles are $1\,$mM.
The insets of each panel are the fitted exponents $\alpha$ as a function of $\phi_\mathrm{w}$.
The symbols are the average of measurements using four different cells, and the error bar denotes the standard deviations.
}
\label{fig:27_1}
\end{figure}

The $-\Im Z$ spectra of sample liquids (open circles in Fig.~\ref{fig:27}) are plotted in Fig.~\ref{fig:49}.
The spectra are the average of measurements using four different cells. 
The lines in different colors denote the water volume fraction $\phi_\mathrm{w}=1.0$ (purple), $0.8$ (blue), $0.6$ (light blue), $0.4$ (green), $0.2$ (orange), and $0.0$ (red), where the spectra of other volume fractions are not plotted because of the visibility. 
All the spectra are similar in shape to that derived from the equivalent circuit model in Fig.~\ref{fig:1}d.
In the range of high frequency, all the spectra have a characteristic upward slope.
The amplitude of these high-frequency slopes is almost linearly increasing from $\phi_\mathrm{w}=0$ to $1$.  
In contrast, in the range of low frequency, all the spectra have a characteristic downward slope.
The equivalent circuit model derives eq.~\ref{eq:imz} for the low-frequency slope, and the amplitudes of these downward slopes reflect the double-layer capacitance.
Although the spectra of the NaCl solutions (Fig.~\ref{fig:49}aei) lack $\phi_\mathrm{w}=0$, they exhibit very similar amplitude of the downward slopes to each other, meaning that the structure of the double layer are not affected by the addition of 1-propanol.
The spectra of other salts (TBAC, NaClO$_4$, and TBAClO$_4$) have similar trends, but the amplitudes of downward slopes of $\phi_\mathrm{w}=0$ (red lines) are significantly larger than the other volume fractions.
This suggests that the double-layer structure of 1-propanol-only solutions ($\phi_\mathrm{w}=0$) is significantly different from the others.
The double-layer structure of the intermediate range of $\phi_\mathrm{w}$ is similar to that of water-only solutions ($\phi_\mathrm{w}=1$).   

To analyze this result more quantitatively, $C_\mathrm{d}$ and $\alpha$ were obtained by fitting eq.~\ref{eq:imz} with all spectra of $-\Im Z$ in the frequency range from $10^1\,$Hz to $10^2\,$Hz.
In the fit, $S=2.0\,$cm$^2$ and $\tau_0=(1/2\pi)\,$s are used.
The reason for using $10^1\,$Hz to $10^2\,$Hz is that some experimental spectra for $\phi_\mathrm{w}=0$ exhibit slope variation in the lower frequency from $10^0$ to $10^1\,$Hz.
We think this effect is out-of-range of the equivalent circuit model denoted in Fig.~\ref{fig:1}a. 
Fig.~\ref{fig:27_1} shows the fitted double-layer capacitance, $C_\mathrm{d}$, as a function of the volume fraction of water.
The double-layer capacitances of NaCl solutions are almost constant with varied $\phi_\mathrm{w}$ because the data for $\phi_\mathrm{w}=0$ is not available as shown in Fig.~\ref{fig:27}. 
TBAC, TBAClO$_4$, and NaClO$_4$ solutions exhibit similar trends, and the capacitance of $\phi_\mathrm{w}\neq0$ is almost the similar to those of $\phi_\mathrm{w}=1$ with the same salt concentration.
Furthermore, the capacitance of 1-propanol-only solutions ($\phi_\mathrm{w}=0$) is smaller than those of solutions with water ($\phi_\mathrm{w}\neq 0$).
The reduction of the double-layer capacitance in 1-propanol-only solutions is explained by the combined effect of the size of the molecules and the interfacial dielectric constant in the Stern layer \cite{Hou_2014}.
In the inset of Fig.~\ref{fig:27_1}, the fitted $\alpha$ is shown. 
The solutions that include water ($\phi_\mathrm{w}\neq 0$) exhibit $\alpha=0.9$, whereas 
for the 1-propanol-only solutions ($\phi_\mathrm{w}=0$) the exponent is slightly smaller than $0.9$.

\begin{figure}[t]
\includegraphics{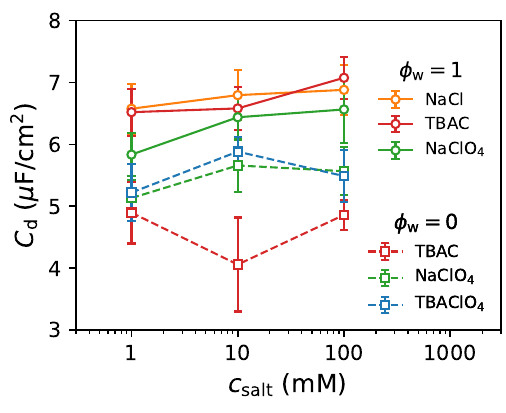}
\caption{
Concentration dependence of the double-layer capacitance for water-only ($\phi_\mathrm{w}=1$, open circles) and 1-propanol-only ($\phi_\mathrm{w}=0$, open squares) solutions.
The symbols are the average of measurements using four different cells, and the error bar denotes the standard deviations.
}
\label{fig:5}
\end{figure}

To see the ion-specific effect in detail, in Fig.~\ref{fig:5}, we plot the double-layer capacitance for water-only ($\phi_\mathrm{w}=1$, open circles) and 1-propanol-only ($\phi_\mathrm{w}=0$, open squares) solutions as a function of the concentrations.  
In both the water-only and 1-propanol-only solutions, the double-layer capacitances are almost constant with varied salinity,  
The constant values are almost independent of the salt types, and it is approximately $6.5\,\mu$F/cm$^2$ for water-only solution whereas $5.0\,\mu$F/cm$^2$ for 1-propanol-only solutions.
This behavior agrees with the recent experimental studies against Stern-Gouy-Chapman theory \cite{Aoki_2013, Khademi_2020}, in which the double-layer capacitance of polycrystal platinum electrode \cite{Aoki_2013} or indium tin oxide electrode \cite{Khademi_2020} exhibits a similar constant capacitance as a function of the salt concentration.  
Although TBAC has a relatively large effect on the double-layer capacitance in 1-propanol-only solutions, 
the experimental results demonstrated that the preferential water adsorption was almost independent of ion species and their concentrations.
Because our system uses aluminum electrodes and 1-propanol as an organic liquid, which are different from platinum electrodes and acetonitrile in the previous study \cite{Aoki_2018}, this result suggests that the preferential water adsorption to electrodes is ubiquitous to broad types of water/organic liquid mixture, metallic electrodes, and electrolytes.
In our study, we expect that the aluminum surfaces are slightly modified to aluminum oxides and hydroxides \cite{Deng_2007}.
This modified layer will be thin enough not to affect the double-layer capacitances, but it is strongly hydrophilic to induce preferential water adsorption.

\section{Conclusions}

We have investigated the double-layer capacitance at the boundary between aluminum electrodes and water/1-propanol electrolyte solutions. 
Four different types of electrolytes and three different concentrations were examined, and the double-layer capacitances were obtained as a function of the volume fraction of water in mixture solvents.
The double-layer capacitances of mixture solvents ($\phi_\mathrm{w}\neq 0$ nor $1$) are almost the same as those of water-only electrolyte solutions ($\phi_\mathrm{w}=1$), and the double-layer capacitances of 1-propanol-only solutions ($\phi_\mathrm{w}=0$) are significantly smaller than those of other water volume fractions.
This demonstrates that the structure of the double-layer of aqueous mixture solvents is similar to those in water-only solutions, and in the water/1-propanol mixture the water molecules are preferentially absorbed to the electrodes. 
This preferential water adsorption to the interface is shown to be ubiquitous because qualitative dependence of the double-layer capacitances on the water volume fraction is independent of the salt types and the salt concentrations.

We make some remarks for future works in this direction as follows.
First, if the electrode surface can be modified to be hydrophobic, the preferential adsorption of organic less-polar molecules will happen. 
In that case, the double-layer capacitances would exhibit opposite variation reported in this study.  
Secondly, some salts destabilize water/organic liquid mixtures into phase separations, which is called the salting-out effect \cite{Gomis_1994}.
It is interesting to observe prewetting transition or voltage-induced surface transition via measurements of the double-layer capacitance. 
We believe this paper will stimulate further studies on the electric double layer in aqueous mixture solvents.

\acknowledgments
This work was supported by JSPS KAKENHI (Grant No. 23K13073), Sasakawa Scientific Research Grant (Grant No. 2021-3001), JST PRESTO (Grant No. JPMJPR21O2).
The authors thank Shunsuke Yabunaka for valuable discussions.

\bibliography{electrochem}


\end{document}